\documentclass[pra,prl,onecolumn,superscriptaddress,floatfix,showpacs,10pt]{revtex4}
\usepackage{amsfonts}
\usepackage{amsmath}
\usepackage{amssymb}
\usepackage{graphicx}

\input{tcilatex}

\begin{document}

\title{\textbf{Thermodynamics Inducing Massive Particles' Tunneling and
Cosmic Censorship }}
\author{Baocheng Zhang}
\affiliation{State Key Laboratory of Magnetic Resonances and Atomic and Molecular
Physics, Wuhan Institute of Physics and Mathematics, Chinese Academy of
Sciences, Wuhan 430071, People's Republic of China}
\affiliation{Graduate University of Chinese Academy of Sciences, Beijing 100081, People's
Republic of China}
\author{Qing-yu Cai}
\affiliation{State Key Laboratory of Magnetic Resonances and Atomic and Molecular
Physics, Wuhan Institute of Physics and Mathematics, Chinese Academy of
Sciences, Wuhan 430071, People's Republic of China}
\author{Ming-sheng Zhan}
\affiliation{State Key Laboratory of Magnetic Resonances and Atomic and Molecular
Physics, Wuhan Institute of Physics and Mathematics, Chinese Academy of
Sciences, Wuhan 430071, People's Republic of China}
\affiliation{Center for Cold Atom Physics, Chinese Academy of Sciences, Wuhan 430071,
China}

\begin{abstract}
By calculating the change of entropy, we prove that the first law of black
hole thermodynamics leads to the tunneling probability of massive particles
through the horizon, including the tunneling probability of massive charged
particles from the Reissner-Nordstr\"{o}m black hole and the Kerr-Newman
black hole. Novelly, we find the trajectories of massive particles are close
to that of massless particles near the horizon, although the trajectories of
massive charged particles may be affected by electromagnetic forces. We show
that Hawking radiation as massive particles tunneling does not lead to
violation of the weak cosmic-censorship conjecture.
\end{abstract}

\maketitle

\section{\textbf{Introduction}}

In 1974, Hawking discovered \cite{swh74}, when considering quantum effects,
that a black hole can emit thermal radiation, which was also confirmed by
many other scientists \cite{hh76,dnp76,wgu76}. Hawking's result does not
coexist with energy conservation, however, since the spacetime background is
fixed in Hawking's calculation \cite{swh75}. Recently, Parikh and Wilczek 
\cite{pw00} developed the method of Hawking radiation as tunneling, in which
the spacetime background was not fixed and the reaction was included. This
method has been discussed in different situations \cite%
{jz05,amv05,jz06,jwc06,zcz09}, and its self-consistency has been checked by
using thermodynamic relations \cite{sk08,bm08,zcz090}.

The laws of thermodynamics play an important role in the area of study on
the black hole. Although the method of Hawking radiation as tunneling can
give the radiation rate, the relation between tunneling dynamics and black
hole thermodynamics still needs to be checked. Recently, Pilling \cite{tp08}
suggested a method of Hawking radiation as tunneling from black hole
thermodynamics. In this method, the change of entropy is calculated to
obtain the tunneling probability directly from the first law of black hole
thermodynamics. The connection between black hole tunneling and
thermodynamics has been verified further \cite{zcz08}, when the tunneling
probability through quantum horizon \cite{amv05,dvf95,dmb02,gm05,dnp05} had
been obtained directly from the first law of black hole thermodynamics \cite%
{zcz08}. However, these discussions only involve the massless particle
tunneling. Since the massless particles do not take charge with themselves,
the trajectory will not be affected by the electromagnetic forces. It is
noted that the tunneling probability of massive charged particles from the
Reissner-Nordstr\"{o}m black hole \cite{jz05} or Kerr-Newman black hole \cite%
{jwc06} has been obtained, but whether the tunneling probability can be
obtained directly from the first law of black hole thermodynamics or not has
not been considered. On the other hand, since the massive particles can take
charge and angular momentum with themselves, the weak cosmic-censorship
conjecture \cite{wi67,dcr74,rmw84} should be revisited with the tunneling of
massive particles. In classical theory, the weak cosmic censorship
conjecture is able to get full support, but the quantum effect may challenge
it \cite{ms07}. So, it is interesting to verify whether the Hawking
radiation as massive particles tunneling could lead to violate the weak
cosmic-censorship conjecture or not.

In this paper, we will calculate the massive particles' tunneling
probability from the first law of thermodynamics and apply this method to
the tunneling from Reissner-Nordstr\"{o}m black hole and Kerr-Newman black
hole. Novelly, we find the trajectories of massive charged particles are
almost identical to those of massless particles near the horizon, although
the trajectories of the massive charged particles connect with the
electromagnetic forces. We will also show that the massive particles
tunneling of a black hole does not violate the weak cosmic-censorship
conjecture, while, in some situations, the Reissner-Nordstr\"{o}m and the
Kerr-Newman black hole may evolve into extreme black holes \cite%
{fh79,hhr95,mp98}.

The paper is organized as follows. In the second section, we calculate the
change of entropy to obtain the probability of massive particle tunneling
from the first law of black hole thermodynamics for a general class of
static, spherically symmetric spacetime. In the third section, we
investigate the massive charged particle tunneling from the Kerr-Newman
black hole and then discuss the situation of Reissner-Nordstr\"{o}m black
hole. Next, we discuss the weak cosmic censorship conjecture. Finally, we
give some discussion and conclusions.

In this paper we take the unit convention $k=\hbar=c=G=1$.

\section{Massive particles' tunneling and thermodynamics}

In this section, we will extend the method by Pilling \cite{tp08} to the
situation of the massive particle tunneling from black hole thermodynamics.
Let us start with the metric

\begin{equation}
ds^{2}=-f\left( r\right) dt_{s}^{2}+\frac{dr^{2}}{g\left( r\right) }%
+r^{2}d\Omega ^{2},  \label{sm}
\end{equation}
where the metric describes a general class of static, spherically symmetric
spacetime. It is noted that there is a coordinate singularity in the metric
at the horizon $r=R$ which is given by $f\left( R\right) =g\left( R\right)
=0 $. For the sake of convenience, we can make the Painlev\'{e} coordinate
transformation to remove the singularity at the horizon,

\begin{equation}
dt\longrightarrow dt_{s}-\sqrt{\frac{1-g\left( r\right) }{f\left( r\right)
g\left( r\right) }}dr,
\end{equation}%
where the time transformation is dependent on $r$, not $t_{{s}}$. So the
metric remains stationary and the time direction is still a Killing vector.
After the transformation, the metric (\ref{sm}) becomes of the following
form:

\begin{equation}
ds^{2}=-f\left( r\right) dt^{2}+2f\left( r\right) \sqrt{\frac{1-g\left(
r\right) }{f\left( r\right) g\left( r\right) }}dtdr+dr^{2}+r^{2}d\Omega ^{2}.
\label{smp}
\end{equation}%
The new metric is called a Painlev\'{e} metric that is regular at the
horizon. The time $t$ here is that of radially free-falling observer through
the horizon.

Generally, for the static coordinates, the coordinate time can represent the
global time due to Einstein's simultaneity. For example, the Schwarzschild
coordinate time in global space can be regarded as the time recorded by a
standard clock that is resting at spatial infinity. The coordinates (\ref%
{smp}) is stationary and not static (the component $g_{{0i}}$ of the metric
tensor does not vanish in the metric (\ref{smp})), so we must check whether
the time is global. According to the general theory \cite{ll75} of the
coordinate clock synchronization in a spacetime, made by Laudau and
Lifshitz, the condition that the simultaneity of coordinate clocks can be
transmitted from one place to another is either the integral $\int \frac{g_{{%
0i}}}{g_{{00}}}dx^{i}=0$ or the integral $\int \frac{g_{{0i}}}{g_{{00}}}%
dx^{i}$ is independent on the path. The condition can be described as
another form \cite{jz05} equivalently,

\begin{equation}
\frac{\partial }{\partial x^{j}}\left( \frac{g_{{0i}}}{g_{{00}}}\right) =%
\frac{\partial }{\partial x^{i}}\left( \frac{g_{{0j}}}{g_{{00}}}\right).
\end{equation}
Obviously, the coordinate (\ref{smp}) satisfies the condition so that the
coordinate time in Painlev\'{e} coordinates can be used as the global time
and the coordinate clock synchronization can be transmitted from one place
to another. This is important for tunneling process in quantum mechanics.

From the coordinate (\ref{smp}), the radial null geodesics can be obtained as

\begin{equation}
\overset{.}{r}=\frac{dr}{dt}=\sqrt{\frac{f\left( r\right) }{g\left( r\right) 
}}\left( \pm 1-\sqrt{1-g\left( r\right) }\right) ,
\end{equation}%
where the positive (negative) sign gives the outgoing (incoming) radial
geodesic, under the implicit assumption that $t$ increases towards the
future.

When considering the massive test particles moving radially in the
background, the radial null geodesics is improper for such particles. In
analogy to Ref. \cite{jz05}, we treat the massive particle as de Broglie
wave and obtain the expression for $\overset{.}{r}$,

\begin{equation}
\overset{.}{r}=v_{{p}}=\frac{1}{2}v_{{g}}=-\frac{1}{2}\frac{g_{{00}}}{g_{{01}%
}}=\frac{1}{2}\sqrt{\frac{f\left( r\right) g\left( r\right) }{1-g\left(
r\right) }},  \label{mg}
\end{equation}%
where $v_{{p}}$ is the phase velocity, and $v_{{g}}$ is the group velocity.
It is obvious that the geodesics of massive particles are different from
radial null geodesics. The surface gravity of the black hole for the
transformed metric (\ref{smp}) at the horizon is given by

\begin{equation}
\kappa _{0}=\Gamma _{{00}}^{0}|_{{r=R}}=\frac{1}{2}\left( \sqrt{\frac{%
1-g\left( r\right) }{f\left( r\right) g\left( r\right) }}g\left( r\right) 
\frac{df\left( r\right) }{dr}\right) |_{{r=R}}.  \label{sg}
\end{equation}%
Generally, when discussing the surface gravity of a black hole, we are
defining a notion that behaves analogously to the Newtonian surface gravity,
while these two things are not the same because the acceleration of a test
body at the event horizon of a black hole turns out to be infinite in
relativity. Generally, the surface gravity of a black hole is not well
defined. However, we can define the surface gravity for a black hole whose
event horizon is a Killing horizon. According to the equation in terms of
time-like Killing $K$\ as $K_{a}\triangledown _{b}K_{a}=\kappa _{0}K_{b}$,
we can obtain the surface gravity in terms of Christoffel components as (\ref%
{sg}). Note that, in static spacetime, a more general expression for black
hole surface gravity should be taken as the geometrical surface gravity \cite%
{hcv09},

\begin{equation}
\kappa =\gamma \kappa _{0}
\end{equation}%
which gives an additional factor comparing with the Eq. (\ref{sg}) and this
factor is a constant at horizon.

Due to $f\left( R\right) =g\left( R\right) =0$, we can expand $f\left(
R\right)$ and $g\left( R\right)$ near the horizon in powers of $r-R$,

\begin{align}
f\left( r\right) & =f^{^{\prime }}\left( R\right) \left( r-R\right) +O\left(
\left( r-R\right) ^{2}\right) ,  \notag \\
g\left( r\right) & =g^{^{\prime }}\left( R\right) \left( r-R\right) +O\left(
\left( r-R\right) ^{2}\right) .  \label{ts}
\end{align}%
According to the laws of black hole thermodynamics, the Hawking temperature
is expressed in terms of the surface gravity via

\begin{equation}
T_{{H}}=\frac{\kappa _{0}}{2\pi }=\frac{\sqrt{f^{^{\prime }}\left( R\right)
g^{^{\prime }}\left( R\right) }}{4\pi }.  \label{rt}
\end{equation}%
Note that the Eq. (\ref{rt}) is not general local temperature expression at
horizon, since it lacks the relative factor stemmed from the use of Kodama
vector instead of the static Killing vector \cite{hcv09}. The general
expression for the temperature can be taken as

\begin{equation}
T=\frac{\kappa }{2\pi }=\frac{\gamma \kappa _{0}}{2\pi }.  \label{grt}
\end{equation}%
Thus we must treat the first law of black hole thermodynamics and the
entropy carefully. The entropy should be given by the area, $A$, of event
horizon as $S=\frac{A}{4\gamma }=\frac{\pi }{\gamma }R^{2}$. Thus the first
law of black hole thermodynamics maintains its form $TdS=dM$.

Let us consider the black hole thermodynamics in the region near the
horizon. When the mass of the black hole changes from $M_{i}$ to $M_{f}$,
the change of the entropy is given as

\begin{equation}
\Delta S=\int dS=\int_{M_{i}}^{M_{f}}\frac{dS}{dM}dM=\int_{M_{i}}^{M_{f}}%
\frac{2\pi }{\gamma }R\frac{dR}{dM}dM.  \label{es}
\end{equation}
Considering the small path near $R$, we can insert the mathematical identity 
$\mathrm{Im}\int_{{r{i}}}^{r{f}}\frac{1}{r-R}dr=-\pi $ in the formula (\ref%
{es}). Thus we obtain

\begin{equation}
\Delta S=-\frac{2}{\gamma }\mathrm{Im}\int_{M_{i}}^{M_{f}}%
\int_{r_{i}}^{r_{f}}\frac{R}{r-R}\frac{dR}{dM}dM.  \label{bhe}
\end{equation}%
With (\ref{grt}) and the expression of the temperature in thermodynamics $%
\frac{1}{T}=\frac{\partial S}{\partial E}$, we can get

\begin{equation}
R\frac{dR}{dM}=\frac{2}{\sqrt{f^{^{\prime }}\left( R\right) g^{^{\prime
}}\left( R\right) }}.  \label{tss}
\end{equation}%
Using the Taylor series (\ref{ts}), we can gain the geodesic (\ref{mg}) of a
massive particle near the horizon,

\begin{equation}
\overset{.}{r}=-\frac{1}{2}\sqrt{f^{^{\prime }}\left( R\right) g^{^{\prime
}}\left( R\right) }\left( r-R\right) +O\left( \left( r-R\right) ^{2}\right) .
\label{rng}
\end{equation}%
It is noticed that the geodesic of massive particle is equivalent to that of
massless particle \cite{tp08} in the first order. That is to say, no matter
whether the particles have mass or not, their trajectories will be close to
each other when they approach the horizon. In particular, their geodesics
coincide precisely with each other at the horizon. One reason for this
phenomenon is that the wavelength of the outgoing particle is infinitely
blueshifted at the horizon. This is also the key argument that the WKB
approximation can be used for Hawking radiation as tunneling. Sometimes, the
infinite blueshift property of the event horizon could lead to the so-called
\textquotedblleft trans-Planckian puzzle\textquotedblright\ \cite{td04},
which is considered as a mathematical artifact of horizon calculations
nowadays \cite{blv05}. On the other hand, the tunneling process is
instantaneous, so the outgoing particles, both the massless and massive
particles, take the same quickest path across the horizon.

With equations (\ref{tss}) and (\ref{rng}), we can obtain the final form of
the change of entropy (\ref{bhe}) as

\begin{equation}
\Delta S=-\frac{2}{\gamma }\mathrm{Im}\int_{M_{i}}^{M_{f}}%
\int_{r_{i}}^{r_{f}}\frac{dR}{\overset{.}{r}}dM=-\frac{2}{\gamma }\mathrm{Im}%
I,
\end{equation}%
where $I$ is the action for an $s$-wave outgoing positive particle in WKB
approximation. Consequently, the tunneling probability is given as

\begin{equation}
\Gamma \thicksim e^{-\frac{2}{\gamma }\mathrm{Im}I}=e^{\Delta S}.
\label{mtp}
\end{equation}%
Thus we obtain the massive particles tunneling probability from the change
of entropy as a direct consequence of the first law of black hole
thermodynamics, along the line presented in Ref. \cite{tp08}. Note that when 
$\gamma =1$, the entropy obeys the Bekenstein-Hawking expression. But the
black hole entropy may have other expressions, for example, when $\gamma =%
\frac{3}{4}$, one has $S=\frac{A}{3}$, which can be obtained from black hole
solution in modified $f(R)$ gravity \cite{bnv04}. This implies that it isn't
always true in general that the related black hole entropy is given by the
Bekenstein-Hawking area law \cite{bnv04}, while the laws of black hole
thermodynamics play an important role in Pilling's argument \cite{tp08}.

\section{Thermodynamics about Kerr-Newman black hole}

The Painlev\'{e} line element of the Kerr-Newman black hole is

\begin{align}
ds^{2}& =-\frac{\Lambda \Sigma }{\left( r^{2}+a^{2}\right) ^{2}-\Lambda
a^{2}\sin ^{2}\theta }dt^{2}+\frac{\Sigma }{r^{2}+a^{2}}dr^{2}+2\frac{\sqrt{%
\left( 2Mr-Q^{2}\right) \left( r^{2}+a^{2}\right) }\Sigma }{\left(
r^{2}+a^{2}\right) ^{2}-\Lambda a^{2}\sin ^{2}\theta }dtdr+\Sigma d\theta
^{2},  \label{knp} \\
A& =\frac{Qr\left( r^{2}+a^{2}\right) }{\left( r^{2}+a^{2}\right)
^{2}-\Lambda a^{2}\sin ^{2}\theta }dt=A_{{t}}dt,
\end{align}%
which is obtained from the Kerr-Newman black hole in the Boyer-Lindquist
coordinate system by the generalized Painlev\'{e}-type coordinate
transformation,

\begin{align}
dt_{{k}}& =dt-\frac{\sqrt{\left( 2Mr-Q^{2}\right) \left( r^{2}+a^{2}\right) }%
}{\Lambda }dr,  \notag \\
d\phi _{{k}}& =d\phi -\frac{a}{\Lambda }\sqrt{\frac{2Mr-Q^{2}}{r^{2}+a^{2}}}%
dr,
\end{align}%
and the dragging coordinate transformation

\begin{equation}
d\phi =\frac{a\left( r^{2}+a^{2}-\Lambda \right) }{\left( r^{2}+a^{2}\right)
^{2}-\Lambda a^{2}\sin ^{2}\theta }dt,
\end{equation}%
where $\Sigma =r^{2}+a^{2}\cos ^{2}\theta $, $\Lambda =r^{2}+a^{2}+Q^{2}-2Mr$%
, $t_{{k}}$ and $\phi _{{k}}$ is the coordinates before transformation. The
form of the vector potential $A$ is unchanged up to a gauge transformation. $%
A_{{t}}$ is the electric potential, and the parameters $M$, $Q$ and $J=Ma$
are the mass, the electric charge, and the angular momentum of the black
hole, respectively. The dragging coordinate transformation is made in order
to make the event horizon coincide with the infinite redshift surface. So
the geometrical optical limit can be applied.

The radial time-like geodesics of massive charged particles are given by

\begin{equation}
\overset{.}{r}=v_{{p}}=\frac{1}{2}v_{{g}}=-\frac{1}{2}\frac{g_{tt}}{g_{tr}}=%
\frac{\Lambda }{2\sqrt{\left( r^{2}+a^{2}\right) \left( r^{2}+a^{2}-\Lambda
\right) }}.
\end{equation}%
Near the outer horizon $r=R^{{+}}=M+\sqrt{M^{2}-Q^{2}-a^{2}}$, which can be
obtained by solving $\Lambda =0$, and the geodesics can be expanded in
powers of $r-R^{+}$ as

\begin{equation}
\overset{.}{r}=\frac{\sqrt{M^{2}-Q^{2}-a^{2}}}{\left( M+\sqrt{%
M^{2}-Q^{2}-a^{2}}\right) ^{2}+a^{2}}\left( r-R^{+}\right) +O\left( \left(
r-R^{+}\right) ^{2}\right) .  \label{kng}
\end{equation}
The radiation temperature is obtained as

\begin{equation}
T_{{H}}=\frac{\kappa }{2\pi }=\frac{1}{2\pi }\frac{\sqrt{M^{2}-Q^{2}-a^{2}}}{%
\left( M+\sqrt{M^{2}-Q^{2}-a^{2}}\right) ^{2}+a^{2}}.  \label{knt}
\end{equation}

Let us consider the black hole thermodynamics in the region near the
horizon. If the energy of the black hole changes from $H_{i}$ to $H_{f}$,
the change of the entropy $S=\pi \left( R^{{+}2}+a^{2}\right) $ is

\begin{equation}
\Delta S=\int_{H_{i}}^{H_{f}}\left( \frac{\partial S}{\partial M}dM+\frac{%
\partial S}{\partial Q}dQ+\frac{\partial S}{\partial J}dJ\right) .
\label{knec}
\end{equation}%
It can be verified that Eq. (\ref{knec}) is another expression of the first
law of black hole thermodynamics, $dM=\frac{\kappa }{2\pi }dS+\Phi dQ+\Omega
dJ$, where $\Phi $ is the electric potential which is defined by $\Phi =A_{{t%
}}=$ $\frac{Qr\left( r^{2}+a^{2}\right) }{\left( r^{2}+a^{2}\right)
^{2}-\Lambda a^{2}\sin ^{2}\theta }$, and $\Omega $ is the dragging angular
velocity which is defined by $\Omega =\frac{d\phi }{dt}=-\frac{g_{{t\phi }}}{%
g_{{\phi \phi }}}=\frac{a\left( r^{2}+a^{2}-\Lambda \right) }{\left(
r^{2}+a^{2}\right) ^{2}-\Lambda a^{2}\sin ^{2}\theta }$. At the horizon, we
have $\Phi ^{{+}}=\Phi |_{{r=R^{+}}}=$ $\frac{Q\left( M+\sqrt{%
M^{2}-Q^{2}-a^{2}}\right) }{\left( M+\sqrt{M^{2}-Q^{2}-a^{2}}\right)
^{2}+a^{2}}$, $\Omega ^{{+}}=\Omega |_{{r=R^{+}}}=\frac{a}{\left( M+\sqrt{%
M^{2}-Q^{2}-a^{2}}\right) ^{2}+a^{2}}$. Considering the conservation of
energy, charge and angular momentum, we have

\begin{align}
\frac{\partial S}{\partial M}& =2\pi R^{{+}}\frac{\partial R^{{+}}}{\partial
M}=\frac{2\pi \left( \left( M+\sqrt{M^{2}-Q^{2}-a^{2}}\right)
^{2}+a^{2}\right) }{\sqrt{M^{2}-Q^{2}-a^{2}}},  \label{kedm} \\
\frac{\partial S}{\partial Q}& =2\pi R^{{+}}\frac{\partial R^{{+}}}{\partial
Q}=\frac{-2\pi Q\left( M+\sqrt{M^{2}-Q^{2}-a^{2}}\right) }{\sqrt{%
M^{2}-Q^{2}-a^{2}}},  \label{kedq} \\
\frac{\partial S}{\partial J}& =2\pi R^{{+}}\frac{\partial R^{{+}}}{\partial
J}=\frac{-2\pi a}{\sqrt{M^{2}-Q^{2}-a^{2}}},  \label{kedj}
\end{align}%
where $a=J/M$. Similarly, these expressions (\ref{kedm}), (\ref{kedq}) and (%
\ref{kedj}) are consistent with the first law of black hole thermodynamics,
that is, $\frac{\partial S}{\partial M}=\frac{2\pi }{\kappa }$, $\frac{%
\partial S}{\partial Q}=\frac{2\pi }{\kappa }\Phi ^{{+}}$, and $\frac{%
\partial S}{\partial J}=\frac{2\pi }{\kappa }\Omega ^{{+}}$. Considering the
small path near the horizon, we can insert the mathematical identity $%
\mathrm{Im}\int_{{r{i}}}^{r{f}}$ $\frac{1}{r-R^{{+}}}dr=-\pi $ in the
formula (\ref{knec}). Thus we obtain

\begin{align}
\Delta S& =-\frac{1}{\pi }\mathrm{Im}\int_{{H{i}}}^{H{f}}\int_{{r{i}}}^{r{f}%
}\left( \frac{1}{r-R^{{+}}}dr\right) \left( \frac{\partial S}{\partial M}dM+%
\frac{\partial S}{\partial Q}dQ+\frac{\partial S}{\partial J}dJ\right)  
\notag \\
& =-2\mathrm{Im}\int_{{H{i}}}^{H{f}}\int_{{r{i}}}^{r{f}}\left( \frac{dr}{%
\kappa \left( r-R^{{+}}\right) }dM-\frac{\Phi ^{{+}}dr}{\kappa \left( r-R^{{+%
}}\right) }dQ-\frac{\Omega ^{{+}}dr}{\kappa \left( r-R^{{+}}\right) }%
dJ\right) ,
\end{align}%
where we have used the expressions of the temperature in thermodynamics $%
\frac{1}{T}=\frac{\partial S}{\partial M}$, the electric potential $\Phi ^{{+%
}}$ and the angular momentum $\Omega ^{{+}}$. Then Eqs. (\ref{kng}) and (\ref%
{kedm})-(\ref{kedj}) give the form of the change of entropy as

\begin{equation}
\Delta S=-2\mathrm{Im}\int_{{H{i}}}^{H{f}}\int_{{r{i}}}^{r{f}}\left( \frac{dr%
}{\overset{.}{r}}dM-\Phi ^{{+}}\frac{dr}{\overset{.}{r}}dQ-\Omega ^{{+}}%
\frac{dr}{\overset{.}{r}}dJ\right)
\end{equation}
Using the Hamilton's equations, $\overset{.}{r}=\frac{dH}{dP_{{r}}}|_{\left(
r;A_{{t}},P_{{A}_{{t}}},\phi ,p_{_{{\phi }}}\right) }=\frac{dM}{dP_{{r}}}$, $%
\overset{.}{A_{{t}}}=\frac{dH}{dP_{{A_{t}}}}|_{\left( A_{{t}};r,P_{{r}},\phi
,p_{_{{\phi }}}\right) }=\Phi ^{{+}}\frac{dQ}{dP_{{A_{t}}}}$ and $\overset{.}%
{\phi }=\frac{dH}{dP_{{\phi }}}|_{\left( \phi ;r,P_{{r}},A_{{t}},P_{{A}_{{t}%
}}\right) }=\Omega ^{{+}}\frac{dJ}{dP_{{\phi }}}$, we can express the change
of entropy in an explicit form as

\begin{equation}
\Delta S=-2\mathrm{Im}\int_{{r{i}}}^{r{f}}\left( P_{{r}}\overset{.}{r}-P_{{A}%
_{{t}}}\overset{.}{A_{{t}}}-P_{{\phi }}\overset{.}{\phi }\right) dt=-2%
\mathrm{Im}I
\end{equation}%
which is related to the emission rate of tunneling particle by $\Gamma \sim
e^{-2\mathrm{Im}I}=e^{\Delta S}$.

Thus we obtain the tunneling probability from the change of entropy of
Kerr-Newman black hole as a direct consequence of the first law of black
hole thermodynamics. It is noticed that when considering the conservation of
energy, charge and angular momentum, the tunneling particle can take the
charge and angular momentum. Note that the generalized coordinate $A_{{t}}$
is an ignorable one and the coordinate $\phi $ is a cyclic one, so the
action can be written as $\int_{{r{i}}}^{r{f}}\left( P_{{r}}\overset{.}{r}%
-P_{{A}_{{t}}}\overset{.}{A_{{t}}}-P_{{\phi }}\overset{.}{\phi }\right) dt$.

In Ref. \cite{jwc06}, it has been pointed out that the tunneling process is
closely related to the first law of black hole thermodynamics, but the
change of angular momentum was treated by the change of mass with the
relation $\delta J=a\delta M$, i.e., the angular momentum of the unit mass
is kept as a constant. In our calculation, although the specific angular
momentum $a$ is not kept as a constant, the first law of black hole
thermodynamics can still lead to the tunneling probability by calculating
the change of entropy. So, our result seems to be more reasonable and more
general. Another potential problem is that of how the angular momentum of
black hole is carried away in the semiclassical WKB approximation, where
only $s$-wave outgoing particles were considered. That is, the potential $%
l\left( l+1\right) /r^{2}$ related to the orbital angular momentum does not
work when calculating the tunneling probability. The angular momentum taken
by the outgoing particles with themselves only means rotation around the
center of the spherically symmetric black hole, not including the
self-spinning. That is the reason why the specific angular momentum is kept
as a constant in Ref. \cite{jwc06}, but our calculation shows that this
restriction is not necessary.

\section{Weak Cosmic Censorship Conjecture}

The uniqueness theorem \cite{wi67,dcr74} states that all stationary black
hole solutions of Einstein-Maxwell equations are uniquely determined by
three conserved parameters: the gravitational mass $M$, the electric charge $%
Q$, and the angular momentum $J$. The three parameters must satisfy the
relation $M^{2}\geqslant Q^{2}+(J/M)^{2}$ to maintain the black hole as the
condition demanded by the weak cosmic-censorship conjecture \cite{rmw84},
which asserts that spacetime singularities coming from the completely
gravitational collapse of a body must be encompassed by the horizon of a
black hole. Particularly, $M^{2}=Q^{2}+(J/M)^{2}$ characterizes extreme
black holes \cite{fh79,hhr95,mp98}, whereas $M^{2}<Q^{2}+(J/M)^{2}$ are
concerned with naked singularities rather than black holes. In what follows
we will check whether the weak cosmic-censorship conjecture will hold or not
in the situations of massive particles' tunneling.

For Kerr-Newman black holes, if the angular momentum of unit mass is
considered as a constant in the tunneling process, the total angular
momentum of black holes is not considered as independent variable and its
change is closely related to the change of mass. Thus at some time the
charge had been taken out of the black hole completely and the mass
continues to decrease until the extremal case $a^{2}=M^{2}$ appears, so the
temperature is absolutely zero and the radiation vanishes. In this case, the
weak cosmic-censorship conjecture will not be violated. If the angular
momentum of unit mass $a$ is not considered as a constant in the tunneling
process, the angular momentum and the charge will be taken out of the black
hole completely before the black hole vanishes because the particles may
take angular momentum with themselves together with that of rotating around
the center of the spherically symmetric black hole. For example, the
electron's $a$ and $Q$ (suitably specified in geometrized units) both exceed
its mass $M$. In such a situation the relation $M^{2}>Q^{2}+a^{2}$ holds
until the black hole vanishes and the extremal black hole will not appear.

Specially, when $a=0$, the coordinate (\ref{knp}) decays into the Painlev\'{e%
} line element of the Reissner-Nordstr\"{o}m black hole. One can easily
approve that the first law of black hole thermodynamics can still give the
massive particles' tunneling rate. The consideration of massive particles'
tunneling is necessary since the massless particles' tunneling may lead to
the violation of weak cosmic-censorship conjecture. On one hand, the
massless particles do not take the charges with themselves but take the
energy by the self-gravity effect when tunneling out of the black hole. The
extreme case $Q^{2}=M^{2}$ can always be reached when the radiation
temperature vanishes. On the other hand, there exist some massless neutral
scalar particles that are spinning. When they tunnel outward, the black hole
will rotate towards the inverse direction due to the conservation of angular
momentum. In this case, no matter how small the angular momentum is, the
relation $M^{2}<Q^{2}+(J/M)^{2}$ can always be reached, because the
particles do not take the charge with themselves. Thus, the weak
cosmic-censorship conjecture will be violated. It is similar to the case of
overspinning a nearly extreme charged black hole via a quantum tunneling
process \cite{ms07}. The violation of weak cosmic censorship conjecture can
be avoided by considering the massive charged particles tunneling. Generally
speaking, the particle charge-mass ratio may be bigger than one; for
example, for the electron $e/m\sim 10^{11}\gg 1$, for the proton $e/m\sim
10^{8}\gg 1$. Thus, the charge will be carried completely out of black hole
soon. The situation of violation of weak cosmic censorship conjecture will
not appear.

In general, the tunneling particles may contain the massless, massive,
massive charged particles in the tunneling process. If the rate of mass loss
is quicker than that of charge and specific angular momentum loss, the
condition $M^{2}=Q^{2}+a^{2}$ may be approached and the extreme black holes
appear as the terminal of the tunneling. If the charge and the angular
moment are always taken out of black holes, the extreme situation will not
appear in the tunneling process. Fortunately, in any case, the weak
cosmic-censorship conjecture will not be violated.

\section{\textbf{Conclusion}}

In summary, the self-consistency of Hawking radiation as massive particles'
tunneling has been verified by the laws of black hole thermodynamics in our
paper. We have showed that the probability of massive particles' tunneling
can be obtained from the first law of thermodynamics by calculating the
change of entropy, and this method has also been used to gain the tunneling
probability of massive charged particles. We have also obtained the
tunneling rate for Kerr-Newman black hole directly from the first law of
black hole thermodynamics when the angular momentum is considered as an
independent variable, i.e., the angular momentum of unit mass $a$ is not
constant in our calculation. Moreover, we have proved that the massive
particles will be along the same trajectory as that of massless particles
when tunneling across the horizon. Finally, we have showed the massive
particles' tunneling does not violate the weak cosmic censorship, while, in
some situations, both a Reissner-Nordstr\"{o}m black hole and a Kerr-Newman
black hole can evolve into extreme black holes.

\section{Acknowledgement}

This work is supported by National Basic Research Program of China (NBRPC)
under Grant No. 2006CB921203.

\end{document}